\documentclass[journal]{IEEEtran}

\usepackage[dvips]{graphicx}
\usepackage{graphicx}
\usepackage{amssymb}
\usepackage{cite}
\usepackage{subfigure}
\usepackage{amsmath}
\usepackage{footnote}

\IEEEoverridecommandlockouts

\begin{document}

\title{Large-Scale Antenna-Assisted Grant-Free Non-Orthogonal Multiple Access via Compressed Sensing}

\author{Jun Fang,~\IEEEmembership{Senior
Member,~IEEE, and Yanlun Wu}
\thanks{Jun Fang and Yanlun Wu are with the National Key Laboratory of Science and Technology on Communications, University of
Electronic Science and Technology of China, Chengdu 611731, China,
Email: JunFang@uestc.edu.cn}
\thanks{This work was supported in part by the National Science Foundation of
China under Grant 61522104.}}

\maketitle

\begin{abstract}
Support massive connectivity is an important requirement in 5G wireless communication system. For massive Machine Type Communication (MTC) scenario, since the network is expected to accommodate a massive number of MTC devices with sparse short message, the multiple access scheme like current LTE uplink would not be suitable. In order to reduce the signaling overhead, we consider an grant-free multiple access system, which requires the receiver facilitate activity detection, channel estimation, and data decoding in ``one shot" and without the knowledge of active user's pilots. However, most of the ``one shot" communication research haven't considered the massive MIMO scenario. In this work we propose a Multiple Measurement Vector (MMV) model based Massive MIMO and exploit the covariance matrix of the measurements to confirm a high activity detection rate.
\end{abstract}

\begin{IEEEkeywords}
5G, MTC, grant-free, Massive MIMO.
\end{IEEEkeywords}

\section{Introduction}
Machine-to-Machine (M2M) is an emerging paradigm for future communication systems. The machine type devices including smart grids, computers, FPGAs, etc, which transmit small data packets occasionally. Due to the sporadic communication and low data rate, these applications put forward new challenges for current LTE Random Access Channel (RACH), where the limited capacity and  excessive control signals are not suitable for massive MTC. One approach for reducing signal overhead in physical layer is to facilitate activity and data jointly detection.

\begin{figure}[!ht]
\centering
\includegraphics[width=2.5in]{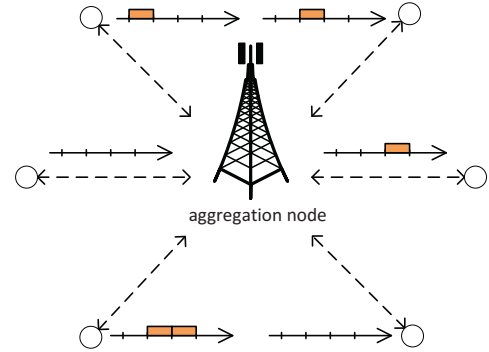}
\caption{sporadic communication scenario.}
\label{fig1}
\end{figure}

In the sporadic transmission scenario, the $K$ sensor nodes are inactive most of the time, as shown in Fig. \eqref{fig1}. Based 5G communication system, \cite{wunder20145gnow} and \cite{wunder2015sparse} introduce a general compressive multiple antenna random access and raised some crucial challenges to physical layer for sporadic communication. Inspired by the sparse user activity, many compressive sensing (CS)-based reliable joint detection both activity and data have been proposed in \cite{Monsees2012Sparsity,bockelmann2013compressive,schepker2013exploiting,
beyene2015compressive,Wunder2014Compressive}. A random channel access facilitated by Code Division Multiple Access (CDMA) has been showed in  \cite{Monsees2012Sparsity,bockelmann2013compressive},  which jointly estimating activity and data based on perfect channel information station (CIS). Pilot sequence based channel estimation has been showed in \cite{schepker2013exploiting,beyene2015compressive,Wunder2014Compressive}, however, the number of nonzero channel coefficients should be known as prior due to the use of Orthogonal Matching Pursuit (OMP) algorithm \cite{schepker2013exploiting,beyene2015compressive}. Based OFDM systems, the base station (BS) with a grant-free transmission procedure has been showed in \cite{Wang2015Compressive,Bayesteh2014Blind}, which also exploit CS for activity detection. Notice that almost all of these methods formulated the detection model as a Single Measurement Vector (SMV) problem which has a poor performance in low SNR. However, in this paper, we making use of the covariance matrix of the measurments, instead of the measurements themselves to achieve a high activity detection rate.

On the other hand, as M2M providing ubiquitous communication environment towards Internet of Things (IoT), the number of M2M users increasing rapidly. But how to increase system capacity? One of the most efficient equipment is massive MIMO system. Massive MIMO is one of the most important technology in 5G system design \cite{wunder2015sparse}, such system can significantly  improve the channel capacity and energy efficiency. The work in \cite{mehmood2013impact} represented that upgrading the BS with Massive MIMO can serve 40 percent more M2M devices in a single cell. Unfortunately, many of the sporadic communication have not considered large-scale antennas at aggregation node. In this paper, we consider a massive MIMO wireless uplink transmission for grant-free non-orthogonal multiple access sporadic communication. In the sporadic traffic scenario, we consider the $K$ sensor nodes communicate with a central aggregation node which we assumed that it equipped with a very large number of antennas, all sensor nodes transmit a small data package occasionally. Our goal is to achieve ``one shot" communication \cite{wunder2015sparse}, i.e., user detection, channel estimation and data detection completed in one time slot. Without any handshaking processing, we will use the sparsity of the active user and non-orthogonal training sequence to avoid excessive signaling overhead. Furthermore, the activity detection is the first and foremost step of ``one shot" communication, a reliable detection is crucial for system performance. According making use of the statistically uncorrelated of Massive MIMO channel information, we show in this paper that the accurate activity detection rate will go to 1 as antennas $M \to \infty$.

\section{SYSTEM MODEL}
\label{sec:system model}

\subsection{Transmission Setup}
We consider a sporadic uplink transmission in TDD massive MIMO system, as shown in figure 1. In order to model the activity of the sensor node, we assume that all sensor nodes have an identical active probability $p_{a}\ll 1$ which is independent of users and data, and the number of active users obeys discrete Binomial distribution $\mathcal B(K, p_a)$ which remain the same in the coherence time. Of course, when the nodes number $K$ increase large, the number of active users is quite small. For an active node $k_{a}$, we assume that it transmits a symbol vector $\mathbf d_{ka} \in \mathcal{A}^N$ per frame, where $\mathcal{A}$ denotes the modulation alphabet. For an inactive node $k_{i}$,  the symbol vector modeled as zero. Consequently, the augmented alphabet $\mathcal{A}_{0}:=\{\mathcal{A} \cup 0\}$ denotes the modulated symbol of both active and inactive nodes.

Different from conventional model in \cite{bockelmann2013compressive}, \cite{schepker2013exploiting}, we assume that the aggregation node equipped with a large number of antennas $M$ which communicates with $K$ sensor nodes each employs with single-antenna (assuming $K\ll M$). Notice that sensor nodes with single-antenna is simple and inexpensive and can be easily extended to nodes with multiple antennas.
The received signal at the aggregation node is can be written as
\begin{align} \label{eq:1}
\mathbf{Y=HX+W}
\end{align}
where $\mathbf{H}\in \mathbb{C}^{M\times K}$ denotes the flat-fading channel matrix, $\mathbf{W}\in \mathbb{C}^{M\times L}$ is the noise matrix with i.i.d entries ${\mathcal{N}(0,\sigma_{w}^2)}_{\mathbb{C}}$. $\mathbf {X}=[\mathbf{x}_1^{T},\mathbf {x}_2^{T},\ldots,\mathbf{x}_K^{T}]^{T}$ and $\mathbf{x}_k\in \mathbb{R}^{L\times1}$, $k=1,\ldots,K$ is the transmitted symbol vector of the $kth$ node and $L$ is the length of symbols. Here, we assume that each node encodes the transmit symbol with a random Gaussian code, owing to the good performance over Gaussian matrices in CS and the other transmission techniques can be seen as a specific case of random coding \cite{schepker2013exploiting}, such as CDMA and SC-FDMA. Further more, we assume that the transmission is synchronized.

In order to implement activity detection, channel estimation and data decoding in ``one shot" \cite{wunder2015sparse}, we split the formulation of \eqref{eq:1} into two separate section. First, for known pilot dictionary, we detect the activity and estimate CIS. Then, after channel estimation, complete data decoding.

\subsection{Massive MIMO Channel Model}\label{channel model}
We assume the aggregation node employs the uniform linear array (ULA) with the antenna spacing $d$ at the aggregation node is half of signal wavelength $\lambda$ (i.e., $d=\frac{1}{2}\lambda$), which has been studied for MIMO channel estimation \cite{rusek2013scaling},\cite{yin2013coordinated},
where the angle spread is limited in the virtual angular domain. Then the channel vector from the $kth$ node to the aggregation node can be written as
\begin{align} \label{eq:2}
\mathbf {h}_{k}=\frac{1}{\sqrt{P}}\sum_{p=1}^{P}g_{kp}a(\theta_{p})
\end{align}
where $P$ is the i.i.d paths from node $k$ to the aggregation node, which related to geometric propagation environment only. $g_{kp}$ is the $kth$ channel's attenuation, which is independent over path direction $p$ and assumed to satisfy $g_{kp}\sim\mathcal{CN}(0,1)$. $a(\theta_{p})$ is the $M\times1$ steering vector which is given by
\begin{align} \label{eq:3}
a(\theta_{p})=[1\ e^{-j2\pi\frac{D}{\lambda}\cos (\theta_{p})}\ \ldots\ e^{-j2\pi\frac{(M-1)D}{\lambda}\cos (\theta_{p})}]^{T}
\end{align}
where $\theta_{p} \in [-\pi/2,\pi/2]$ is a random angle of arrival (AOA).
For massive MIMO system with large arrays, recent research in channel measurement has shown that the channel characteristics approximate the favorable propagation condition fairly well \cite{rusek2013scaling},\cite{yin2013coordinated}, which show that as long as there are enough paths in \eqref{eq:2}, all the channel vectors can be approximated by Gaussian model:
\begin{align} \label{eq:4}
\mathbf h_ k=\mathbf {R}_k^{1/2}\mathbf h_{\mathrm Wk}, k=1,2,\ldots,K,
\end{align}
where $\mathbf h_{\mathrm Wk}\sim\mathbb{C}(\mathbf 0,\mathbf{I}_M)$ and $\mathbf R_k^{1/2}\triangleq \mathbb E\{\mathbf H_k\mathbf H_k^H\}$. Under the most favorable propagation, the elements of channel vector that different users to the same antenna $\mathbf h_m$  can be approximated by independent random variables with zero means. In the rest of this paper, we will hold the assumption of favorable propagation that
\begin{align} \label{eq:H}\nonumber
&E\lbrack\mathbf{h}_m(i)\rbrack = 0\\
&E\lbrack\mathbf{h}_m(i)\mathbf{h}_n(j)\rbrack = \sigma_i^2\delta(i-j)(m-n),\quad 1<m,n\leq M.
\end{align}

\section{Activity Detection}
In this section, we consider the activity and channel information are jointly estimated. In conventional, channel estimation employs orthogonal training sequence, i.e. $L\geq K$ where $L$ represents the length of pilot symbols. As $K$ increases, the training overhead is unaffordable. Furthermore, we consider the grant-free communication where the cumbersome handshaking procedure is no longer applicable for 5G. As the study in \cite{schepker2013exploiting}, the communications from senor nodes to aggregation node are always sporadic, i.e., sensor nodes transmit signals occasionally. By the theorem of compressed sensing, the sparsity of sporadic communications can be leveraged to reduce the overhead of nodes accessing (i.e. $L<K$) and a grant-free communication can be achieved. Note that the training sequences used for sensor nodes in a slot are not allocated by the aggregation node. However, the aggregation node knows the codebook of training symbols.

In the stage of activity detection, the received pilot signal at aggregation node is
\begin{align} \label{channel init}
\mathbf{Y}_p=\mathbf{HS}^H+\mathbf{W}
\end{align}
taking a simply transposition of \eqref{channel init}, we can get
\begin{align} \label{eq:MMV}
\mathbf{Y}^{H}_p=\mathbf{SH}^{H}+\mathbf{W}^H
\end{align}
where $\mathbf{W}^H\in \mathbb{C}^{L\times M}$ is the noise matrix with i.i.d entries ${\mathcal{N}(0,\sigma_{w}^2)}_{\mathbb{C}}$ and $\mathbf{S}=[\mathbf {s}_1,\mathbf {s}_2,\ldots,\mathbf {s}_K]$ is the $L\times K$ training dictionary with $L$ is the length of the pilot sequences. Note that $L<K$, the dictionary is underdetermined. $\mathbf{H}^H=[\mathbf{h}_1,\mathbf{h}_2,\ldots,\mathbf{h}_M]$ is a $K\times M$ matrix, where $\mathbf{h}_m$ is a sparse vector
\begin{displaymath}\label{eq:h}
\mathbf{h}_m = \left\{ \begin{array}{ll}
h_m^{(k)} & \textrm{$k \in S_0$}\\
0 & \textrm{elsewhere}
\end{array} \right.
\end{displaymath}
and they share a common sparsity $\Vert \mathbf{h}_m\Vert_{0}=D$, where $D$ denotes the sparse level of the active nodes, $S_0=\{i_1,i_2,\ldots,i_D\}$ denotes the index of nonzeros. Consequently, $\mathbf{Y}^H$ contains $M$ measurement vectors which share a common sparse support, this is a MMV problem in sparse representation, it has been well discussed and showed a good performance in \cite{chen2006theoretical,wipf2007empirical,zhang2011sparse}.

\subsection{Activity Detection Model}
In the section of activity detection, our goal is to recovery the sparse support of $\mathbf{H}$ without estimate the underlying data, after detecting the active nodes correctly, channel estimation can be done with the least squares (LS) approach.

The activity detection is the first step of "one shot" communication and it is crucial for system performance. To avoid data losing caused by fallacious classification, we consider the detection based on covariance matrix of the measurement to reduce the interference of noise, and the statistic characteristics of noise variance usually can be seen as a prior. Based on MMV model in \eqref{eq:MMV}, let $\mathbf{y}_m^{H}\in\mathbb{C}^{L\times 1}, m=1,\ldots,M$ denote a set of $M$ antennas' measurement vectors, and $\mathbf{h}_m^{H}\in\mathbb{C}^{K\times 1}, m=1,\ldots,M$ are $M$ unknown channel vectors with the uniform sparsity $D$, we have
\begin{align} \label{eq:6}
\mathbf{y}^{H}_m=\mathbf{S}\mathbf{h}_m^{H} + \mathbf{w}_m^{H},\quad 1\leq m \leq M.
\end{align}

We are interested in the correlation matrix of received signal, which is given by
\begin{align}\label{Ey^H*y}
E(\mathbf{y}_m^{H}\mathbf{y}_m)=E(\mathbf{S}\mathbf{h}_m^H\mathbf{h}_m\mathbf{S}^H)
+E(\mathbf{w}_m^H\mathbf{w}_m)
\end{align}
Denote the sample covariance matrix as $\mathbf{\Phi}_{yy}$, we can get a simplified formulation as
\begin{align}\label{eq:correlation}
\mathbf{\Phi}_{yy} \triangleq \frac{1}{M}\sum_{m=1}^{M}\mathbf{y}_m^{H}\mathbf{y}_m=\mathbf{S}\mathbf{R}_{hh}\mathbf{S}^H + \mathbf{\Phi}_{ww}
\end{align}
where $\mathbf{R}_{hh}=\frac{1}{M}\sum_{m=1}^{M}\mathbf{h}_m^{H}\mathbf{h}_m$. With $M\to \infty$, exploiting \eqref{eq:H}, we can get that $\mathbf{R}_{hh}$ goes to a diagonal matrix with $r_{kk} \neq 0$ corresponding an active node and $r_{kk} = 0$ if the $k$th node is inactive, so the activity detection fully determined by the nonzero position of the matrix $\mathbf{R}_{hh}$. $\mathbf{\Phi}_{ww}=\frac{1}{M}\sum_{m=1}^{M}\mathbf{w}_m^{H}\mathbf{w}_m$ is a matrix obeys the Wishart distribution and can be written as $\mathbf{\Phi}_{ww} = \sigma_w^2\mathbf{I}_L + \mathbf{N}$, where $\sigma_w^2\mathbf{I}_L$ denotes the mean of the noise covariance matrix and $\mathbf{N}$ is a random variable with zeros mean and variance $\sigma_w^4/M$,  so $\mathbf{N}$ converge to zero with $M \to \infty$.

Using the property of Khatri-Rao product, upon vectorization of \eqref{eq:correlation}, we have
\begin{align} \label{eq:8}
\mathbf{z} \triangleq vec(\mathbf{\Phi}_{yy})=(\mathbf{S}\odot\mathbf{S})\mathbf{r}_{hh}+\mathbf{n}+\mathbf{e}
\end{align}
\begin{align} \label{eq:9}
\mathbf{x} \triangleq \mathbf{z}-\mathbf{n}=(\mathbf{S}\odot\mathbf{S})\mathbf{r}_{hh}+\mathbf{e}
\end{align}
where $\mathbf{n}=vec(\mathbf{\Phi}_{ww})$ is a $L^2\times 1$ vector. With the assumption that the noise variance is known, then the activity detection can be reformulated as \eqref{eq:9} by finding the support of $\mathbf{r}_{hh}\in \mathbb{C}^{K\times 1}$. $\mathbf{r}_{hh}$ is a $D$ sparse vector with non-zero elements comprised in the vector $diag(\mathbf{R}_{hh})$, whose non-zero elements are given by
\begin{align} \label{eq:DiagRhh}
\sigma^{2[M]}_i \triangleq \frac{1}{M}\sum_{m=1}^{M}[\mathbf{h}_m^{H}]^2_i, \qquad i \in S_0
\end{align}

Owing to the finite number of antennas $M$, an additive noise-like term $\mathbf{e}\in \mathbb{C}^{L^2\times 1}$ arising which can be divided into two parts $\mathbf{e=e_1+e_2}$ , the elements of $\mathbf{e_1}$ and $\mathbf{e_2}$ are given respectively

\begin{align} \label{eq:e1}
 \frac{1}{M}\sum_{i\neq j}\sum_{m=1}^M\mathbf{S}_{p,i}\mathbf{S}_{q,j}[\mathbf{h}_m^{H}]_i[\mathbf{h}_m^{H}]_j, \ 1\leq p, q \leq L, \ 1\leq i, j \leq K.
\end{align}

\begin{align}\label{eq:e2}
\frac{1}{M}\sum_{p\neq q}\sum_{m=1}^M[\mathbf{w}_m^{H}]_p[\mathbf{w}_m^{H}]_q, \ 1\leq p, q \leq L.
\end{align}
They reflect the cross-correlation of $\mathbf y_m$ for finite number of antennas. This noise-like term will goes to arbitrarily small when the number of antennas $M\to \infty$.

To approach the activity detection, we have convert the MMV model in \eqref{eq:MMV} to SMV model in \eqref{eq:9} by recovering the sparse vector $\mathbf{r}_{hh}$, where $\mathbf{S}\odot\mathbf{S}$ is the efficient measurement matrix with dimension $L^2\times K$, as we show shortly, the dimension gain in measurement matrix could efficiently improve the system capacity of active nodes.

We focus on recover the support of sparse vector $\mathbf r_{hh}$, note that there is a constraint for vector $\mathbf r_{hh}$ to be non-negative, we consider the constraint $l_1$ norm regularized quadratic programming:
\begin{align} \label{eq:LASSO}\nonumber
&\min_{r}\bigg( \frac{1}{2}\Vert(\mathbf{S}\odot\mathbf{S})\mathbf{r}-\mathbf{x}^{[M]}\Vert^2_2+\lambda \|r\|_1\bigg)\\
&\text{s.t.}\quad \mathbf{r}\geq 0
\end{align}
where $\lambda$ a regularization parameter. This problem, known as LASSO \cite{tibshirani1996regression}, whose performance has been studied in \cite{wainwright2009sharp}, \cite{schmidt2005least} including the sufficient conditions for exact sparsity recovery and examined a variety of approaches proposed for parameter estimation in LASSO.

\subsection{Theoretical Results}
In this section, we will give some theoretical results of training code design and activity detection. For ease of notation, we assume the columns of training dictionary $\mathbf{S}$ to have $\|\mathbf{s}_i\|_2=1$. Notice that this assumption has no impact on the recovering performance. The mutual coherence of matrix $\mathbf{S}$ is $\mu_{\mathbf S}$, which is defined as:
\begin{align}\label{eq:u}
\mu_{\mathbf S}\triangleq \max_{i \neq j}|\mathbf{s}_i^H\mathbf{s}_j|
\end{align}

Let $\mu_{\mathbf S_{KR}}$ indicates the mutual coherence of $\mathbf S\odot\mathbf S$, combine the property of Kronecker products that $(\mathbf a_i\otimes\mathbf b_i)^H(\mathbf a_j\otimes\mathbf b_j)=\big(\mathbf a_i^H\mathbf a_j\big)\big(\mathbf b_i^H\mathbf b_j\big)$, we can obtain
\begin{align}
\big|[\mathbf s_{KR}]_i^H[\mathbf s_{KR}]_j\big|=\big|(\mathbf s_i\otimes\mathbf s_i)^H(\mathbf s_j\otimes\mathbf s_j)\big|=\big(|\mathbf s_i^H\mathbf s_j|\big)^2
\end{align}
Hence the connection between $\mu_{\mathbf S_{KR}}$ and $\mu_{\mathbf S}$ can be written as:
\begin{align}\label{eq:mu_KR}
\mu_{\mathbf S_{KR}}=\mu_{\mathbf S}^2
\end{align}

Based on mutual coherence of a matrix, many recovery performance in different situations have been studied \cite{Ben2010Coherence,Donoho2006Stable,Tropp2006Corrigendum}. Follows from the study of \cite{Donoho2003Optimally}, solving the sparse support problem via $\min_{\mathbf x,\mathbf {Sx=y}}\Vert\mathbf x\Vert_1$ yields a sufficient condition based on mutual coherence that
\begin{align}\label{condtion}
|S_0|<\frac{1}{2}\big(1+\frac{1}{\mu_{\mathbf S}}\big)
\end{align}
 Consequently, through the mutual covariance matrix and Khatri-Rao product, the coherence of measurement matrix from $\mu_{\mathbf S}$ becomes $\mu_{\mathbf S}^2$, owing to the fact that $\mu_{\mathbf S}<1$, the capacity of system active nodes $|S_0|$ has been raised.

 Further more, the support recovery condition in \eqref{condtion} suffers from the Welch bound \cite{welch1974lower}, which is defined as:
\begin{align}\label{eq:welch}
\mu_S \geq \sqrt{\frac{K-L}{(K-1)L}}.
\end{align}

If the maximum number of active node $|S_0|$ has determinant,
we can easily calculate the lower bound for the needed length of pilot sequence from \eqref{eq:mu_KR}, \eqref{condtion}, and \eqref{eq:welch}, which is given by
\begin{align}\label{eq:15}
L > \frac{2K|S_0|-K}{K+2|S_0|-2}
\end{align}

Note that there are many construction matrix nearly meeting the Welch bound \eqref{eq:welch}, such as the Gold/Gold-like codes, FZC codes, Kasami codes, etc. In sporadic communication system, via choosing a proper codec we can obtain the shortest length of pilot to ensure active nodes communication and meanwhile improve the efficiency of data transmission in a coherence time.

%
In the following analysis, we will show that the probability of active node detection will great than $1-\alpha\beta^{-M}$, which increase overwhelming with the number of antennas $M$.

According to the study of \cite{pal2015pushing} (\emph{Theorem 7}), if $|S_0|<\frac{1}{2}\Big(1+\frac{1}{\mu^2_S}\Big)$, the optimal solution $\mathbf{r^*}$ to the constrained LASSO is unique and it satisfies $Supp(r^*)=S_0$ when the following two conditions hold:
\begin{align}\label{eq:E12}
\|\mathbf{e}\|_2<\lambda\frac{1+\mu_S^2-2\mu_S^2D}{1+\mu_S^2-\mu_S^2D}, \qquad \textrm{(Event \textit{E1})}\\
\sigma_{\textrm{min}}^{2[M]}>\frac{\|\mathbf{e}\|_2+\lambda}{1+\mu_S^2-\mu_S^2D}, \quad \textrm{(Event \textit{E2})}
\end{align}
where $\sigma_{\textrm{min}}^{2[M]}=\textrm{min}\{\sigma_i^{2[M]},i\in S_0\}$.

We first state the probability of successful recovery of sparse by solving the LASSO \cite{pal2015pushing}, which can be directly used in our derivation:
\newtheorem{theorem}{Theorem}
\begin{theorem}
The probability of successful recovery of sparse support by solving the LASSO, denotes as $\textit{P}_s$, satisfies:
\begin{align}\nonumber
\textit{P}_s&\geq\mathcal{P}\left(\emph{E1}\cap\emph{E2}\right)\\\nonumber
&=\mathcal{P}\Big(\|\mathbf{e}\|_2<\lambda\frac{1+\mu_S^2-2\mu_S^2D}{1+\mu_S^2-\mu_S^2D},\nonumber
\sigma_{\textrm{min}}^{2[M]}>\frac{\|\mathbf{e}\|_2+\lambda}{1+\mu_S^2-\mu_S^2D}\Big)\\\nonumber
&\geq\mathcal{P}\Big(\|\mathbf{e}\|_2<c_1,\sigma_{\textrm{min}}^{2[M]}>c_2\Big)\\
&\geq\prod_{i=1}^{D}\mathcal{P}(\sigma_i^{2[M]}>c_2)-\sum_{i=1}^{L^2}\mathcal{P}\Big(|[\mathbf{e}]_i|
\geq\frac{c_1}{L}\Big)
\end{align}
where
\begin{align}
c_1\triangleq\lambda\frac{1+\mu_S^2-2\mu_S^2D}{1+\mu_S^2-\mu_S^2D}, \ c_2\triangleq\lambda\frac{2(1+\mu_S^2)-3\mu_S^2D}{\big(1+\mu_S^2-\mu_S^2D\big)^2}
\end{align}
\end{theorem}

There is an important inequality in our derivation which is shown in\cite{haupt2010toeplitz}:

\emph{Lemma} 1: Let each of $x_i$ and $y_i, i=1,\ldots,k$ be uncorrelated zero mean Gaussian random variables with variance $\sigma_x^2$ and $\sigma_y^2$ respectively. Then
\begin{align}\nonumber
\mathcal{P}\left(\bigg|\sum_{i=1}^kx_iy_i\bigg|\geq t\right)\leq2\textrm{exp}\left(-\frac{t^2}{2\sigma_x\sigma_y(2\sigma_x\sigma_yk+t)}\right)
\end{align}

Notice from \eqref{eq:e1}, we know that
$\big|[\mathbf e_1]_i\big|\leq\frac{\|\mathbf{S}\|_{\infty,\infty}^2}{M}
\sum_{i\neq j}\sum_{m=1}^{M}\Big|[\mathbf {h}_m^H]_i[\mathbf h_m^H]_j\Big|$, if $\big|[\mathbf e_1]_i\big|\geq C_1$, then

\begin{align}\nonumber
&\sum_{i\neq j}\sum_{m=1}^{M}\big|[\mathbf {h}_m^H]_i[\mathbf h_m^H]_j\big|\geq\frac{C_1M}{\|\mathbf{S}\|_{\infty,\infty}^2} \Longrightarrow\\\label{C1}
&\sum_{i_0\neq j_0}\big|[\mathbf {h}_m^H]_{i_0}[\mathbf{h}_m^H]_{j_0}\big|
>\frac{C_1M}{\|\mathbf{S}\|_{\infty,\infty}^2D(D-1)}
\end{align}

where $i_0,j_0\in{1,\ldots,D},i_0\neq j_0, D=|S_0|$. According to \eqref{C1}, we can get that
\begin{align}\label{P_c1}\nonumber
\mathcal{P}&\big(|[\mathbf e_1]_i|\geq C_1\big)\\
&\leq\mathcal{P}\Big(\sum_{m=1}^{M}\big|[\mathbf {h}_m^H]_{i_0}[\mathbf{h}_m^H]_{j_0}\big|>\frac{C_1M}{\|\mathbf{S}\|_{\infty,\infty}^2D(D-1)}\Big)
\end{align}

Under the assumption that $[\mathbf {h}_m^H]_{i_0}$ and $[\mathbf{h}_m^H]_{j_0}$ are uncorrelated zero mean Gaussian random variables, we can use Lemma 1 to obtain
\begin{align}\nonumber
\mathcal{P}&\Big(\sum_{m=1}^{M}\Big|[\mathbf{h}_m^H]_{i_0}[\mathbf{h}_m^H]_{j_0}
\Big|>\frac{C_1M}{\|\mathbf{S}\|_{\infty,\infty}^2D(D-1)}\Big)\\\nonumber
&\leq2\textrm{exp}\left(-\frac{Mt_1^2}{2\sigma_{i_0}\sigma_{j_0}(2\sigma_{i_0}\sigma_{j_0}+t)}\right)
\\\label{ineq:e1}
&\leq2\textrm{exp}\bigg(-\frac{Mt_1^2}{2\sigma_{\textrm{max}}^{(1)}\sigma_{\textrm{max}}^{(2)}
(2\sigma_{\textrm{max}}^{(1)}\sigma_{\textrm{max}}^{(2)}+t_1)}\bigg)
\end{align}
where $t_1=\frac{C_1M}{\|\mathbf{S}\|_{\infty,\infty}^2D(D-1)}$, $\sigma_{\textrm{max}}^{(k)}$ denotes the $k$th largest element in the set of $\{\sigma_i\}_{i=1}^{|S_0|}$.

Similarly, from \eqref{eq:e2} and Lemma 1, we can say that
\begin{align}\nonumber
\mathcal{P}&\big(|[\mathbf e_2]_i|\geq C_2\big)\\\nonumber
&\leq\mathcal{P}\Big(\sum_{m=1}^{M}\big|[\mathbf {w}_m^H]_{p}[\mathbf{w}_m^H]_{q}\big|>\frac{C_2M}{L(L-1)}\Big)\\\nonumber
&\leq2\textrm{exp}\bigg(-\frac{Mt_2^2}{2[\sigma_\textrm{w}]_p[\sigma_\textrm{w}]_q
(2[\sigma_\textrm{w}]_p[\sigma_\textrm{w}]_q+t_2)}\bigg)\\\label{ineq:e2}
&\leq2\textrm{exp}\bigg(-\frac{Mt_2^2}{2[\sigma_\textrm{w}]_{\textrm{max}}^{(1)}
[\sigma_\textrm{w}]_{\textrm{max}}^{(2)}
(2[\sigma_\textrm{w}]_{\textrm{max}}^{(1)}[\sigma_\textrm{w}]_{\textrm{max}}^{(2)}+t_2)}\bigg)
\end{align}
where $t_2=\frac{C_2M}{L(L-1)}$, $[\sigma_\textrm{w}]_{\textrm{max}}^{(k)}$ denotes the $k$th largest element in the set of $\{[\sigma_w]_i\}_{i=1}^{L}$.

Now, we assume that $C_1+C_2=\frac{c_1}{L}$, notice that $|[\mathbf e_1]_i+[\mathbf e_2]_i|\leq|[\mathbf e_1]_i|+|[\mathbf e_2]_i|$ and we can get $\mathcal P(\big|[\mathbf e_1]_i + [\mathbf e_2]_i\big|\geq\frac{c_1}{L})\leq\mathcal P(|[\mathbf e_1]_i|+|[\mathbf e_2]_i|\geq\frac{c_1}{L})$. Besides, $\mathcal P(|[\mathbf e_1]_i|\leq C_1)\mathcal P(|[\mathbf e_1]_i|\leq C_1)\leq\mathcal P(|[\mathbf e_1]_i|+|[\mathbf e_2]_i|\leq C_1+C_2)$, then we can obtain
\begin{align}\nonumber
\mathcal{P}\big(|\mathbf e_i|\geq\frac{c_1}{L}\big)=&\mathcal{P}\big(|[\mathbf e_1]_i+[\mathbf e_2]_i|\geq\frac{c_1}{L}\big)\\\nonumber
\leq&\mathcal{P}\big(|[\mathbf e_1]_i|+|[\mathbf e_2]_i|\geq\frac{c_1}{L}\big)\\\nonumber
=&\mathcal{P}\big(|[\mathbf e_1]_i|+|[\mathbf e_2]_i|\geq C_1+C_2)\\\nonumber
=&1-\mathcal{P}\big(|[\mathbf e_1]_i|+|[\mathbf e_2]_i|< C_1+C_2)\\\nonumber
\leq&1-\mathcal{P}\big(|[\mathbf e_1]_i|< C_1)\mathcal{P}\big(|[\mathbf e_2]_i|< C_2)\\\nonumber
=&1-\big(1-\mathcal P(|[\mathbf e_1]_i|\geq C_1)\big)\big(1-\mathcal P(|[\mathbf e_2]_i|\geq C_2)\big)\\\nonumber
=&\mathcal P(|[\mathbf e_1]_i|\geq C_1)+\mathcal P(|[\mathbf e_2]_i|\geq C_2)-\\
&\mathcal P(|[\mathbf e_1]_i|\geq C_1)\mathcal P(|[\mathbf e_2]_i|\geq C_2)
\end{align}

Define that $\delta_1\triangleq\Big(\frac{t_1^2}{2\sigma_{\textrm{max}}^{(1)}\sigma_{\textrm{max}}^{(2)}
(2\sigma_{\textrm{max}}^{(1)}\sigma_{\textrm{max}}^{(2)}+t_1)}\Big)$ and  $\delta_2\triangleq\Big(\frac{t_2^2}{2[\sigma_\textrm{w}]_{\textrm{max}}^{(1)}[\sigma_\textrm{w}]_{\textrm{max}}^{(2)}
(2[\sigma_\textrm{w}]_{\textrm{max}}^{(1)}[\sigma_\textrm{w}]_{\textrm{max}}^{(2)}+t_2)}\Big)$, combing with \eqref{ineq:e1} and \eqref{ineq:e2}, we get
\begin{align}\label{eq:term2}\nonumber
\sum_{i=1}^{L^2}\mathcal{P}&\big(|\mathbf e_i|\geq\frac{c_1}{L}\big)\\\nonumber
\leq&L^2\Big[\mathcal P(|[\mathbf e_1]_i|\geq C_1)+\mathcal P(|[\mathbf e_2]_i|\geq C_2)-\\\nonumber
&\mathcal P(|[\mathbf e_1]_i|\geq C_1)\mathcal P(|[\mathbf e_2]_i|\geq C_2)\Big]\\
\leq&2L^2\Big(e^{-\delta_1M}+e^{-\delta_2M}-e^{-(\delta_1+\delta_2)M}\Big)
\end{align}

\emph{Lemma} 2: Let $x_i, i=1,\ldots, M$ denote independent zero mean Gaussian random variables with variance $\sigma_i^2$. Assume $0<C<\sigma_{\textrm{min}}^2$, then, there exist $\beta>1$ such that $\mathcal P\big(\frac{1}{M}\sum_{i=1}^Mx_i^2>C\big)\geq1-\beta^{-M}$.

\emph{Proof} 2: Denote $p_x \triangleq \frac{1}{M}\sum_{i=1}^{M}x_i^2$. Then $\mathcal P\big(p_x>C\big)=\mathcal P\Big(\sum_{i=1}^{M}x_i^2>CM\Big)=\mathcal P\Big(\sum_{i=1}^Mz_i^2>\frac{CM}{\sigma_i^2}\Big)$ where $z_i$ denote i.i.d zero mean standard normal variables. Therefore $\sum_{i=1}^{M}z_i^2$ is a Chi-Squared random variable with $M$ degrees of freedom. Observe
\begin{align}\label{beta}\nonumber
\mathcal P(p_x>C)&=1-\mathcal P\Big(\sum_{i=1}^{M}z_i^2\leq\frac{CM}{\sigma_i^2}\Big)\\\nonumber
&\geq1-\textrm{exp}\Big(\frac{tCM}{\sigma_i^2}(1+2t)^{-M/2}\Big)\\
&\geq1-\textrm{exp}\Big(\frac{tCM}{\sigma_{\textrm{min}}^2}(1+2t)^{-M/2}\Big)
\end{align}
where $t>0$. This equation is based on Chernoff Bound and Moment-generating function of a Chi-Squared that random variable with $M$ degrees of freedom is given by $(1-2t)^{-M/2}, t<1/2$. Now define the function $\beta(t)\triangleq\textrm{ exp}\big(-\frac{2tC}{\sigma_{\textrm{min}}^2}\big)(1+2t)$
we can write from \eqref{beta} that
\begin{align}\label{Beta}
\mathcal P(p_x>C)\geq1-\big(\beta(t)\big)^{-M/2}
\end{align}
We want to ensure that $\exists t>0$, such that $\beta(t)>1$. Notice that if $\beta\triangleq\textrm{ exp}\big(-\frac{2tC}{\sigma_{\textrm{min}}^2}\big)(1+2t)>1$, we will get $C<\gamma(t)$, where $\gamma(t)\triangleq\frac{\sigma_{\textrm{min}}^2}{2t}\textrm{log}(1+2t)$.
We can easily find that $\gamma(t)$ is a decreasing function in $t$ for $t>0$. Using the limit theorem of equivalent
 infinitesimal we can get that $\gamma(0)=\sigma_{\textrm{min}}^2$. Given that $C<\sigma_{\textrm{min}}^2$,
then indeed $\exists t_0>0$ such that $C<\gamma(t_0)$
which in turn implies $\beta(t_0)>1$. Hence, we conclude that $\mathcal P\Big(\frac{1}{M}\sum_{i=1}^{M}x_i^2>C\Big)\geq1-\beta^{-M}$ for $\beta=\sqrt{\beta(t_0)}>1$.

Then from \eqref{Beta}, we can say that
\begin{align}\label{eq:term1}
\prod_{i=1}^{|S_0|}\mathcal{P}(\sigma_i^{2[M]}>c_2)\geq\prod_{i=1}^{|S_0|}(1-\beta_i^{-M})
\end{align}
where $\sigma_i^{2[M]}$ is defined in \eqref{eq:DiagRhh} and $\beta_i>1$.

\emph{Theorem} 2:  The probability of successful recovery of sparse support by solving the LASSO \eqref{eq:LASSO} is greater than $1-\alpha\gamma^{-M}$ with $\gamma>1$.

\emph{Proof}: From \eqref{eq:term2} and \eqref{eq:term1}, we can rewrite the recovery probability $\mathcal P_s$ as
\begin{align}\label{eq:Ps}\nonumber
\mathcal P_s&\geq\prod_{i=1}^{D}\mathcal P\big(\sigma_i^{2[M]}>c_2\big)-\sum_{i=1}^{L^2}\mathcal P\Big(|[\mathbf e]_i|\geq\frac{c_1}{M}\Big)\\\nonumber
&\geq\prod_{i=1}^{|S_0|}(1-\beta_i^{-M})-2L^2(e^{-\delta_1M}+e^{-\delta_2M}-e^{-(\delta_1+\delta_2)M})\\\nonumber
&\geq1-\sum_{i=1}^{|S_0|}\beta_i^{-M}-2L^2(e^{-\delta_1M}+e^{-\delta_2M}-e^{-(\delta_1+\delta_2)M})\\
&\geq1-D\beta_{\textrm{min}}^{-M}-2L^2(e^{-\delta_1M}+e^{-\delta_2M}-e^{-(\delta_1+\delta_2)M})
\end{align}
 notice that $\delta_1>0, \delta_2>0$. Define $1<\gamma<\textrm{min}(\beta_{\textrm{min}},e^{\delta_1},e^{\delta_2})$, then
\begin{align}\label{eq:result}\nonumber
\mathcal P_s&\geq1-D\gamma^{-M}-4L^2\gamma^{-M}+2L^2e^{-(\delta_1+\delta_2)M}\\
&\geq1-(D+4L^2)\gamma^{-M}
\end{align}

Note that when the aggregation node with large antenna arrays (i.e., $M \to \infty$), the recover probability goes to 1 with sparsity level restricted upto $\frac{1}{2}\big(1+\frac{1}{\mu_S^2}\big)$, this will greatly improves the performance of present sporadic communication which have not been considered in massive MIMO system.

\section{Channel Estimation and Data Decoding via LS}
After estimating the activity correctly, aggregation node will get a new $L \times K_a$ matrix $\mathbf{\hat{S}}$, which contains $K_a$ active nodes training sequences occupation in current slot. Denote $\mathbf{\hat H}$ as a new $M\times K_a$ matrix which contains the support vectors only. Then we can rewrite the $M\times L$ received signal as
\begin{align}\label{eq:hat}
\mathbf{Y}_p=\mathbf{\hat H \hat S}^H+\mathbf{W}
\end{align}

To simplify system design, we implement channel estimation $\mathbf{\hat H}$ combining with received pilot signal $\mathbf{Y}_p$ and known training matrix $\mathbf{\hat{S}}$ according to Least Squares(LS) approach
\begin{align}\label{eq:16}
\mathbf{\hat{H}} = \mathbf{Y}_p\mathbf{\hat{S}}^\dag =  \mathbf{Y}_p \mathbf{\hat{S}}\big(\mathbf{\hat{S}}^H \mathbf{\hat{S}}\big)^{-1}
\end{align}

After channel estimation, the transmit data can be decoded through received signal at aggregation node. The received signal matrix  can be written as
 \begin{align}\label{eq:17}
\mathbf{Y}_d=\mathbf{\hat{H}D}+\mathbf{W}
\end{align}
where $\mathbf{D}=[\mathbf d_1, \mathbf d_2,\ldots, \mathbf d_N]$ is a ${K_a\times N}$ matrix comprises $N$ modulation symbols of $K_a$ active sensor nodes. Also, the symbol vector for an inactive node $k_i$ is modeled as zeros rather than modulation symbols over one slot.

For known channel $\mathbf{\hat{H}}$, a LS approach for data decoding can be obtained from \eqref{eq:17}
\begin{align}\label{eq:18}
\mathbf{\hat{D}} = \mathbf{\hat{H}}^\dag\mathbf{Y}_d = \mathbf{\hat{H}}^H\big(\mathbf{\hat{H}} \mathbf{\hat{H}}^H\big)^{-1}\mathbf{Y}_d.
\end{align}

\section{Simulation Results}
In the simulations we consider $K=64$ users, all the users are synchronized to the aggregation synchronization signal. Here we suppose the coefficients of each row in matrix $\mathbf H$ are independent random variables with zero means as described in Section \ref{channel model}. We consider a random Gaussian code which is normalized for pilot symbols.

\subsection{Performance of Activity Detection}
As described in Section \ref{sec:system model}, firstly, we should complete activity detection according to \eqref{eq:LASSO} using training dictionary. We compare the activity detection performance of our covariance matrix  method with three traditional MMV algorithms:
\begin{enumerate}
\item MSBL \cite{wipf2007empirical}: An extension of Sparse Bayesian Learning (SBL) for the SMV model to the MMV model. For a fairly comparison, we set the true noise variance as the noise variance parameter value and freeze it.
\item BOMP \cite{fang2012block}: As mentioned at \cite{zhang2011sparse}, the MMV model can be transformed to a block SMV model. By letting $\mathbf {y}=\mathrm{vec}(\mathbf{Y}_p)\in\mathbb{C}^{LM\times 1}$, $\mathbf {T}=\mathbf{S}\otimes\mathbf{I}_M$, $\mathbf {h}=\mathrm{vec}(\mathbf{H})\in\mathbb{C}^{KM\times 1}$, $\mathbf {w}=\mathrm{vec}(\mathbf{W})$, we can transform the MMV model to block SMV model: $\mathbf {y=Th+w}$, which can be solved by block-OMP (BOMP).
\item MFOCUSS \cite{cotter2005sparse}: The regularized M-FOCUSS is developed for noisy data. We set the p-norm to 0.8 as suggested by the authors.
\end{enumerate}

In comparison to the activity detection error rates of these different approaches, we investigate a system with the number of antennas $\mathrm M=128$ at aggregation node. For channel matrix $\mathbf H$, the $|S_0|$ non-zero columns were randomly selected in each trial. For training matrix $\mathbf S$, we use a length-20 random code for each pilot symbol. Since that the length of pilot sequence is less to the number of users $K$, the training matrix is non-orthogonal. The probability estimates are computed by 1000 Monte Carlo Runs and a successful detection is defined as recovering the entire true support.

\begin{figure}[!h]
\centering
\includegraphics[width=7cm] {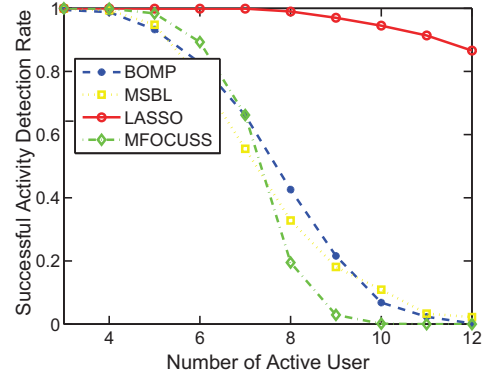}
\caption{Percentage of successful active user detection, as a function of $|S_0|$. Here L=20, M=128, SNR=0.}
\label{fig2}
\end{figure}

\begin{figure}[!h]
\centering
\includegraphics[width=7cm] {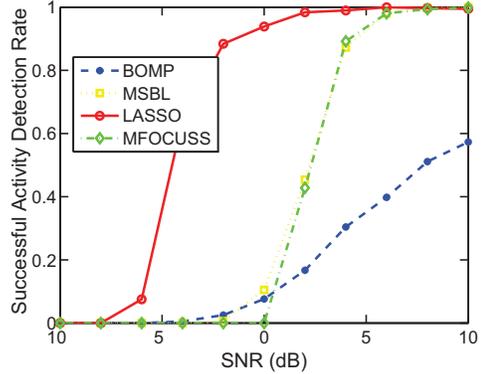}
\caption{Percentage of successful active user detection, as a function of SNR. Here L=20, M=128, $|S_0|$=10.}
\label{fig3}
\end{figure}

\begin{figure}[!h]
\centering
\includegraphics[width=7cm] {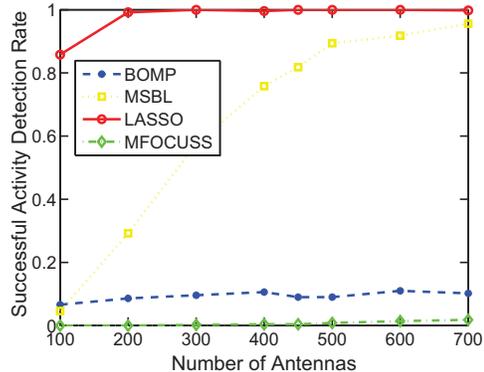}
\caption{Percentage of successful active user detection, as a function of M. Here L=20, $|S_0|$=10, SNR=0.}
\label{fig4}
\end{figure}

In Fig. \ref{fig2}, We plot the activity detection performance as a function of active users $|S_0|$. The results show that for less than 4 active users, all the algorithms mentioned above have an almost perfect detection rate, but as the number of active users increase over 6, only the LASSO based on covariance matrix method remain a high detection rate. Consider the same experimental setup as before and fixed $|S_0|=10$, assume an active user transmits with an average energy per symbol of $E_S=1$, then the signal to noise ratio (SNR) is $1/\sigma_w^2$, Fig. \ref{fig3} shows the activity detection rate as a function of SNR from -10 dB to 10 dB. It can be seen that there is a distinct advantage of LASSO at low SNR, at almost 0dB, LASSO based covariance matrix has reached a very high detection rate. In Fig. \ref{fig4}, we plot the detection rate as a function of the antennas number $M$ with $|S_0|=10$. The observation is that given this level of sparsity and the particular noise, the detection rate of MSBL and LASSO increase as $M$ becomes large, but MFOCUSS and BOMP are almost incapable of recovering the entire true active users at all values of $M$. Owing to this LASSO method based on covariance matrix, there are a perfect detection rate as the number of antennas equipped at BS becomes large. It demonstrates that this method is very satisfied with the massive MIMO system.

\begin{figure}[!h]
\centering
\includegraphics[width=7cm] {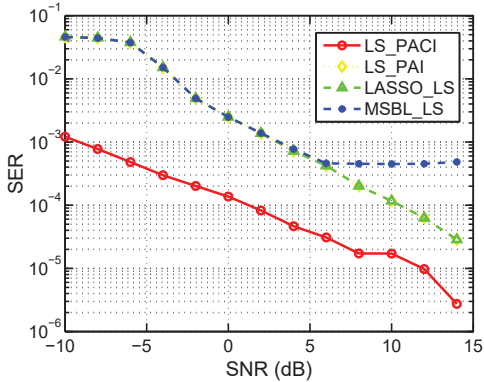}
\caption{Symbol error as a function of SNR. Here L=20, T=40,M=500, $|S_0|$=6.}
\label{fig5}
\end{figure}

\begin{figure}[!h]
\centering
\includegraphics[width=7cm] {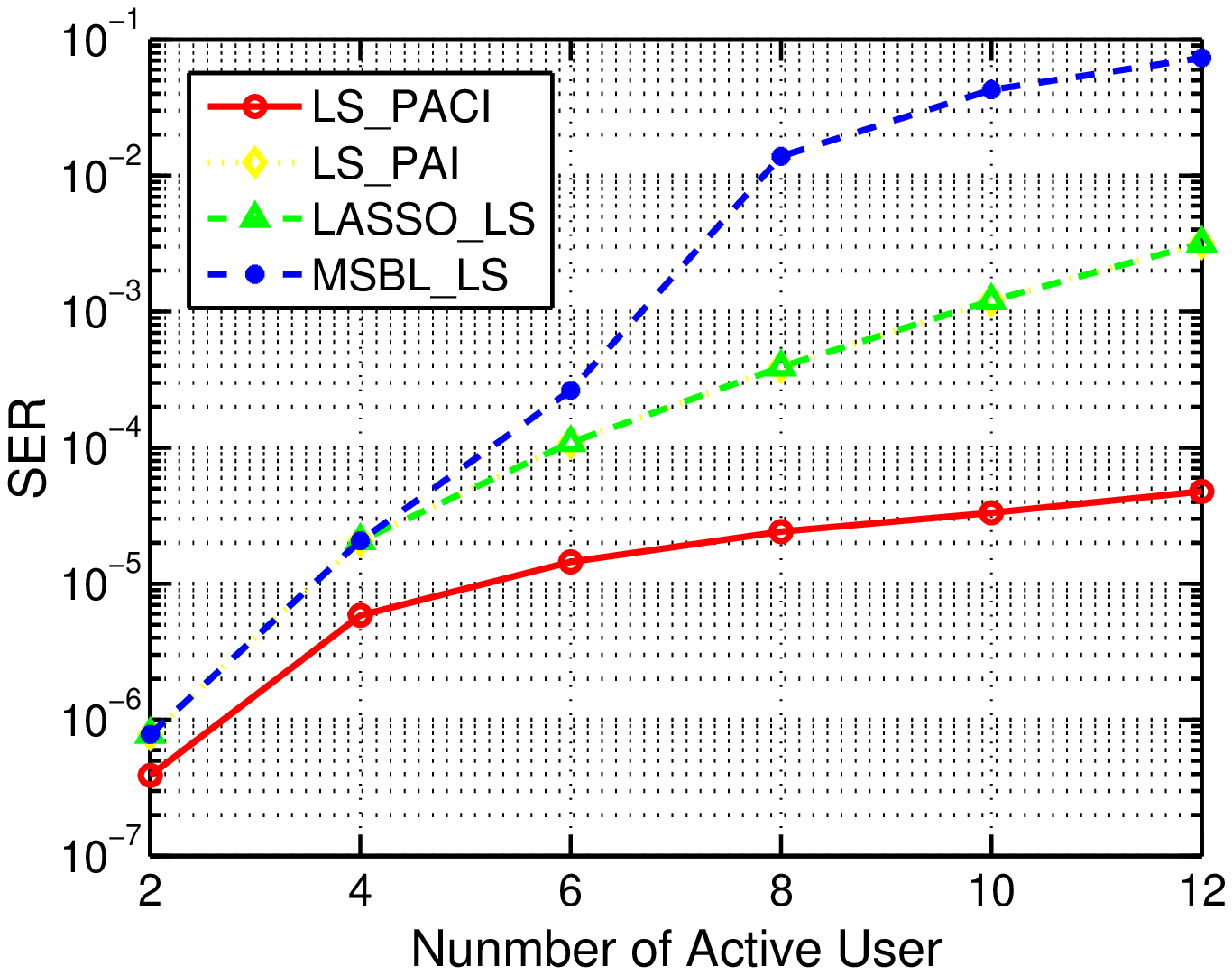}
\caption{Symbol error as a function of $|S_0|$. Here L=20, T=40,M=500, SNR=10.}
\label{fig6}
\end{figure}

\begin{figure}[!h]
\centering
\includegraphics[width=7cm] {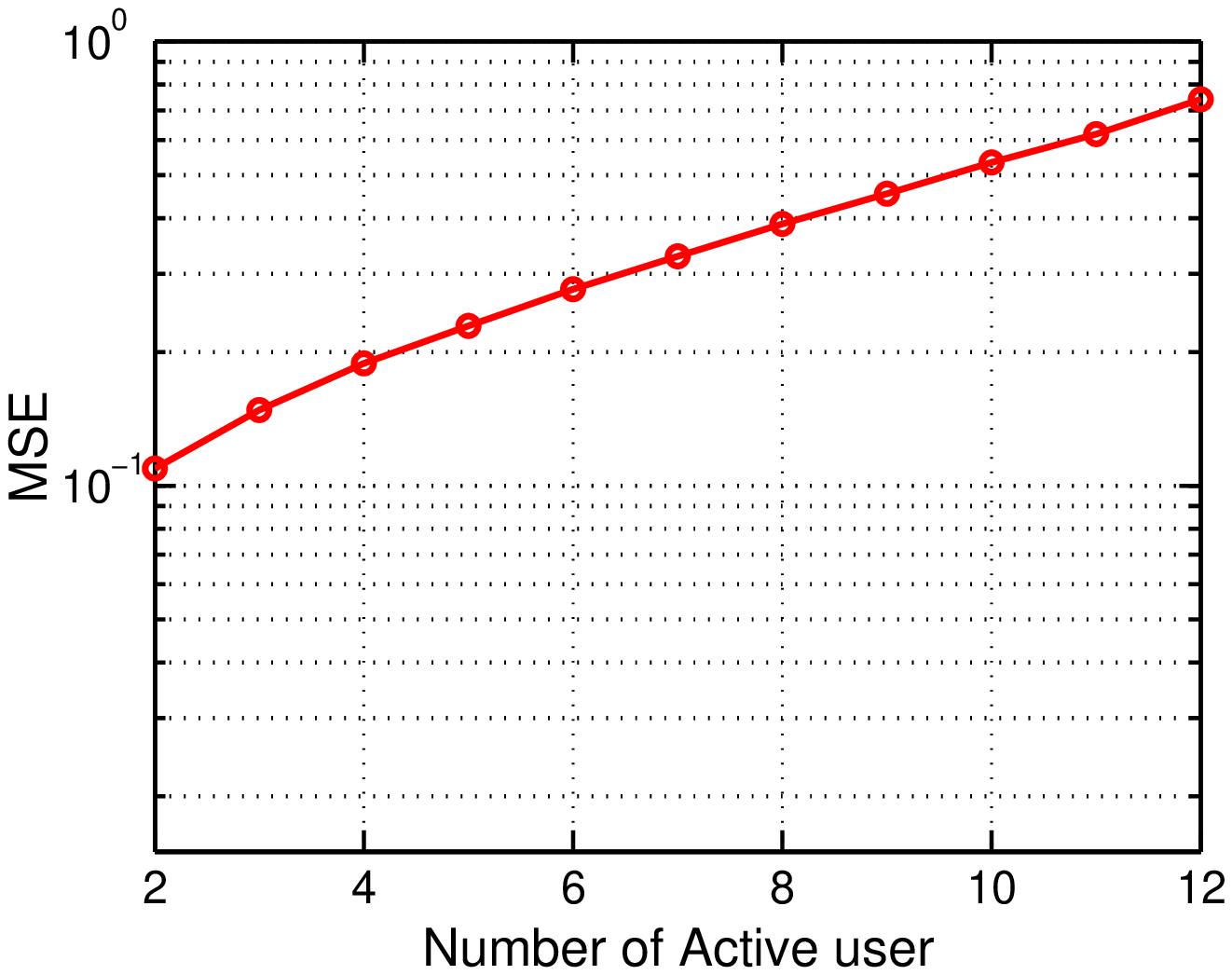}
\caption{Channel estimation error, as a function of $|S_0|$. Here L=20, M=500, SNR=10.}
\label{fig7}
\end{figure}

\begin{figure}[!h]
\centering
\includegraphics[width=7cm] {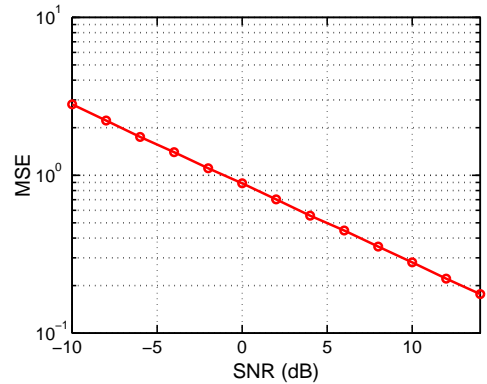}
\caption{Channel estimation error, as a function of SNR. Here L=20, M=500, $|S_0|$=6.}
\label{fig8}
\end{figure}
In Fig. \ref{fig5} and Fig. \ref{fig6}, we simulated the average Symbol Error Rate (SER) over the augmented alphabet $\mathcal A$ for a length of $N=40$ transmit-symbols, the symbols are also encoded by random coding. Due to the data decoding is the second stage of ``one shot" communication, the performance of SER is the result of missed detection and channel estimation error. In order to study the contribution of the channel and activity estimation error, we measured the SER for perfect activity and channel information (PACI) through LS and perfect activity information (PAI) through LS for channel estimation and data decoding, which are compared with the activity detection based LASSO and MSBL. The simulation results indicate that the performance of SER with and without PAI doesn't has significant differences in low activity and high SNR, with the increase of noise and active users, LASSO based activity detection shows certain advantages. We can also observe that there is a large gap between PACI and other cases in SER. This is mainly due to the LS based channel estimation with non-orthogonal training matrix suffers significantly, so the performance of SER is mainly limited by channel estimation. In Figs. \ref{fig7} and Figs. \ref{fig8}, we plot the Mean Squared Error to explore the performance of channel estimation based LS:
\begin{align}\label{Channel MSE}
\mathcal{MSE} =\sum_{ka\in|S_0|}
\frac{\big\|\mathbf{h}_{ka}-\mathbf{\hat{h}}_{ka}\big\|_2^2}{\big\| \mathbf{h}_{ka}\big\|_2^2}
\end{align}


\section{Conclusion} \label{sec:conclusion}
In this paper, we have proposed a massive MIMO wireless uplink transmission based grant-free non-orthogonal
multiple access for 5G. Numerical results show that our proposed model achieves a
significant performance improvement through exploiting statistical information about
the unknown massive MIMO channel information.

 \bibliography{newbib}

\begin{thebibliography}{10}
\providecommand{\url}[1]{#1}
\csname url@samestyle\endcsname
\providecommand{\newblock}{\relax}
\providecommand{\bibinfo}[2]{#2}
\providecommand{\BIBentrySTDinterwordspacing}{\spaceskip=0pt\relax}
\providecommand{\BIBentryALTinterwordstretchfactor}{4}
\providecommand{\BIBentryALTinterwordspacing}{\spaceskip=\fontdimen2\font plus
\BIBentryALTinterwordstretchfactor\fontdimen3\font minus
  \fontdimen4\font\relax}
\providecommand{\BIBforeignlanguage}[2]{{%
\expandafter\ifx\csname l@#1\endcsname\relax
\typeout{** WARNING: IEEEtran.bst: No hyphenation pattern has been}%
\typeout{** loaded for the language `#1'. Using the pattern for}%
\typeout{** the default language instead.}%
\else
\language=\csname l@#1\endcsname
\fi
#2}}
\providecommand{\BIBdecl}{\relax}
\BIBdecl

\bibitem{wunder20145gnow}
G.~Wunder, P.~Jung, M.~Kasparick, T.~Wild, F.~Schaich, Y.~Chen, S.~Brink,
  I.~Gaspar, N.~Michailow, A.~Festag \emph{et~al.}, ``5gnow: non-orthogonal,
  asynchronous waveforms for future mobile applications,'' \emph{Communications
  Magazine, IEEE}, vol.~52, no.~2, pp. 97--105, 2014.

\bibitem{wunder2015sparse}
G.~Wunder, H.~Boche, T.~Strohmer, and P.~Jung, ``Sparse signal processing
  concepts for efficient 5g system design,'' \emph{Access, IEEE}, vol.~3, pp.
  195--208, 2015.

\bibitem{Monsees2012Sparsity}
F.~Monsees, C.~Bockelmann, D.~Wubben, and A.~Dekorsy, ``Sparsity aware
  multiuser detection for machine to machine communication,'' in \emph{GLOBECOM
  Workshops}, 2012, pp. 1706 -- 1711.

\bibitem{bockelmann2013compressive}
C.~Bockelmann, H.~F. Schepker, and A.~Dekorsy, ``Compressive sensing based
  multi-user detection for machine-to-machine communication,''
  \emph{Transactions on Emerging Telecommunications Technologies}, vol.~24,
  no.~4, pp. 389--400, 2013.

\bibitem{schepker2013exploiting}
H.~F. Schepker, C.~Bockelmann, and A.~Dekorsy, ``Exploiting sparsity in channel
  and data estimation for sporadic multi-user communication,'' in
  \emph{Wireless Communication Systems (ISWCS 2013), Proceedings of the Tenth
  International Symposium on}.\hskip 1em plus 0.5em minus 0.4em\relax VDE,
  2013, pp. 1--5.

\bibitem{beyene2015compressive}
Y.~Beyene, C.~Boyd, K.~Ruttik, C.~Bockelmann, O.~Tirkkonen, and R.~Jantti,
  ``Compressive sensing for mtc in new lte uplink multi-user random access
  channel,'' in \emph{AFRICON, 2015}.\hskip 1em plus 0.5em minus 0.4em\relax
  IEEE, 2015, pp. 1--5.

\bibitem{Wunder2014Compressive}
G.~Wunder, P.~Jung, and C.~Wang, ``Compressive random access for post-lte
  systems,'' in \emph{IEEE International Conference on Communications
  Workshops}, 2014, pp. 539--544.

\bibitem{Wang2015Compressive}
B.~Wang, L.~Dai, Y.~Yuan, and Z.~Wang, ``Compressive sensing based multi-user
  detection for uplink grant-free non-orthogonal multiple access,'' in
  \emph{Vehicular Technology Conference}, 2015.

\bibitem{Bayesteh2014Blind}
A.~Bayesteh, E.~Yi, H.~Nikopour, and H.~Baligh, ``Blind detection of scma for
  uplink grant-free multiple-access,'' in \emph{International Symposium on
  Wireless Communications Systems}, 2014, pp. 853--857.

\bibitem{mehmood2013impact}
Y.~Mehmood, N.~Haider, W.~Afzal, U.~Younas, I.~Rashid, and M.~Imran, ``Impact
  of massive mimo systems on future m2m communication,'' in
  \emph{Communications (MICC), 2013 IEEE Malaysia International Conference
  on}.\hskip 1em plus 0.5em minus 0.4em\relax IEEE, 2013, pp. 534--537.

\bibitem{rusek2013scaling}
F.~Rusek, D.~Persson, B.~K. Lau, E.~G. Larsson, T.~L. Marzetta, O.~Edfors, and
  F.~Tufvesson, ``Scaling up mimo: Opportunities and challenges with very large
  arrays,'' \emph{Signal Processing Magazine, IEEE}, vol.~30, no.~1, pp.
  40--60, 2013.

\bibitem{yin2013coordinated}
H.~Yin, D.~Gesbert, M.~Filippou, and Y.~Liu, ``A coordinated approach to
  channel estimation in large-scale multiple-antenna systems,'' \emph{Selected
  Areas in Communications, IEEE Journal on}, vol.~31, no.~2, pp. 264--273,
  2013.

\bibitem{chen2006theoretical}
J.~Chen and X.~Huo, ``Theoretical results on sparse representations of
  multiple-measurement vectors,'' \emph{IEEE Transactions on Signal
  Processing}, vol.~54, no.~12, pp. 4634--4643, 2006.

\bibitem{wipf2007empirical}
D.~P. Wipf and B.~D. Rao, ``An empirical bayesian strategy for solving the
  simultaneous sparse approximation problem,'' \emph{Signal Processing, IEEE
  Transactions on}, vol.~55, no.~7, pp. 3704--3716, 2007.

\bibitem{zhang2011sparse}
Z.~Zhang and B.~D. Rao, ``Sparse signal recovery with temporally correlated
  source vectors using sparse bayesian learning,'' \emph{Selected Topics in
  Signal Processing, IEEE Journal of}, vol.~5, no.~5, pp. 912--926, 2011.

\bibitem{tibshirani1996regression}
R.~Tibshirani, ``Regression shrinkage and selection via the lasso,''
  \emph{Journal of the Royal Statistical Society. Series B (Methodological)},
  pp. 267--288, 1996.

\bibitem{wainwright2009sharp}
M.~J. Wainwright, ``Sharp thresholds for high-dimensional and noisy sparsity
  recovery using-constrained quadratic programming (lasso),'' \emph{Information
  Theory, IEEE Transactions on}, vol.~55, no.~5, pp. 2183--2202, 2009.

\bibitem{schmidt2005least}
M.~Schmidt, ``Least squares optimization with l1-norm regularization,''
  \emph{CS542B Project Report}, pp. 14--18, 2005.

\bibitem{Ben2010Coherence}
Z.~Ben-Haim, Y.~C. Eldar, and M.~Elad, ``Coherence-based performance guarantees
  for estimating a sparse vector under random noise,'' \emph{IEEE Transactions
  on Signal Processing}, vol.~58, no.~10, pp. 5030--5043, 2010.

\bibitem{Donoho2006Stable}
D.~L. Donoho, M.~Elad, and V.~N. Temlyakov, ``Stable recovery of sparse
  overcomplete representations in the presence of noise,'' \emph{IEEE
  Transactions on Information Theory}, vol.~52, no.~1, pp. 6--18, 2006.

\bibitem{Tropp2006Corrigendum}
J.~A. Tropp, ``Corrigendum in "just relax: Convex programming methods for
  identifying sparse signals in noise",'' \emph{IEEE Transactions on
  Information Theory}, vol.~52, no.~3, pp. 1030--1051, 2006.

\bibitem{Donoho2003Optimally}
D.~L. Donoho and M.~Elad, ``Optimally sparse representation in general
  (nonorthogonal) dictionaries via ?'' \emph{Proceedings of the National
  Academy of Sciences}, vol. 100, no.~5, pp. 2197--202, 2003.

\bibitem{welch1974lower}
L.~R. Welch, ``Lower bounds on the maximum cross correlation of signals
  (corresp.),'' \emph{Information Theory, IEEE Transactions on}, vol.~20,
  no.~3, pp. 397--399, 1974.

\bibitem{pal2015pushing}
P.~Pal and P.~Vaidyanathan, ``Pushing the limits of sparse support recovery
  using correlation information,'' \emph{Signal Processing, IEEE Transactions
  on}, vol.~63, no.~3, pp. 711--726, 2015.

\bibitem{haupt2010toeplitz}
J.~Haupt, W.~U. Bajwa, G.~Raz, and R.~Nowak, ``Toeplitz compressed sensing
  matrices with applications to sparse channel estimation,'' \emph{IEEE
  Transactions on Information Theory}, vol.~56, no.~11, pp. 5862--5875, 2010.

\bibitem{fang2012block}
J.~Fang and H.~Li, ``Block-sparsity pattern recovery from noisy observations,''
  in \emph{Acoustics, Speech and Signal Processing (ICASSP), 2012 IEEE
  International Conference on}.\hskip 1em plus 0.5em minus 0.4em\relax IEEE,
  2012, pp. 3321--3324.

\bibitem{cotter2005sparse}
S.~F. Cotter, B.~D. Rao, K.~Engan, and K.~Kreutz-Delgado, ``Sparse solutions to
  linear inverse problems with multiple measurement vectors,'' \emph{Signal
  Processing, IEEE Transactions on}, vol.~53, no.~7, pp. 2477--2488, 2005.

\end{thebibliography}
\bibliographystyle{IEEEtran}

\end{document}